\documentclass{elsart}
\usepackage{graphics,psfig,epsfig,amsmath,amssymb,latexsym,subfigure}

\begin{document}
\runauthor{Makatchev and Kozlov}
\begin{frontmatter}
%INCLUDE NEW PAPER FROM IC.AC.UK
\title{On the Cell-based Complexity of\\
Recognition of Bounded Configurations by\\
Finite Dynamic Cellular Automata}

\author{Maxim Makatchev}
%\author[MSU]{Vadim N. Kozlov}

\address{Learning Research and Development Center, University of Pittsburgh,\\
Pittsburgh, PA 15260, USA (maxim@pitt.edu)}

%\address[MSU]{Dept. of Mathematical Theory of Intelligent Systems,
%Faculty of Mechanics and Mathematics,
%Lomonosov Moscow State University, Moscow, Russia (mk@cs.msu.su)}

%\maketitle

\begin{abstract}
This paper studies complexity of recognition of classes of bounded configurations by 
a generalization of conventional cellular automata (CA) --- finite dynamic cellular automata (FDCA). 
Inspired by the CA-based models of biological and computer vision, this study attempts to derive the properties of a complexity measure and of the classes of input configurations that make it beneficial to realize the recognition via a two-layered automaton as compared to a one-layered automaton. A formalized model of an image pattern recognition task is utilized to demonstrate that the derived conditions can be satisfied for a non-empty set of practical problems.
\end{abstract}
\begin{keyword}
Finite dynamic cellular automata; Complexity measures; Image pattern recognition; Biologically-motivated computer vision 
\end{keyword}

\end{frontmatter}

\section{Introduction}

Multi-scale image processing techniques have been finding natural implementations in hierarchical pyramid/neural networks \cite{spence}, cellular neural networks \cite{CY,fajfar,KCY,orovas}, CA-based models \cite{hernandez,Makatchev1998,sahota}, including probabilistic cellular automata \cite{brady}. Cellular automata have also been adapted to serve as a model of certain functions of primary visual cortex \cite{Makatchev1997}. The fact that such applications of CA are concerned with input configurations of bounded size limits the usefulness of conventional CA complexity measures that do not take into account properties of individual cells (e.~g. size of the basic neighborhood). For example, an arbitrary class of input configurations of bounded size can be recognized (in the sense defined in section~\ref{sec:recognition}) by a cellular automaton in a single step of the computation by means of using a basic neighborhood of a large enough size to fully cover the area of the input configurations. Obviously, this implementation is not the one preferred in computer vision and biological vision systems. Instead, multilayered designs are utilized in a wide range of image processing algorithms  \cite{rosenfeld}, as well as in biological vision systems \cite{Hubel}. What properties of the complexity measures of the visual systems result in efficiency of the multilayered approach? What properties of bounded configuration classes make them efficiently recognizable by multilayered CA? These are the two questions we address in this paper.

 One immediate explanation of a potentially faster performance of a multilayered structure is its straightforward parallelization guaranteed by functional independence of the cells of the same layer. This paper, however, shows that the speedup can be achieved even with a sequential model of computation.
% ..
% If the possibility of the straightforward parallelization of an algorithm implementing a multilayered structure would be the only way to achieve a speed-up over an algorithm implementing the recognition via a single layer, presumably one would not bother with multilayered structures when targeting a sequential computer (at least if the running time was the decisive criterion). Obvi ...run faster their implementation on sequential computers presumably would not give .

One of the complications of adapting a conventional CA to a real-life recognition application arises from the fact that the cells of a conventional CA have a fixed functionality at every step of the computation. While this property of CA is suitable for representation of hardware arrays of processing units \cite{khan,tzionas,uhr}, this is an artificial constraint for software implementations and for modelling of biological systems where neuron layers of different functionalities realize step-wise processing. To overcome this constraint, in section~\ref{sec:FDCA} we define a generalization of CA, a dynamic cellular automaton (DCA) and its finite version, a finite dynamic cellular automaton (FDCA). DCA is different from the structurally dynamic cellular automata introduced in \cite{SDCA} in that the local map and basic neighborhood of DCA are preprogrammed for each step of the computation, as opposed to being derived on-the-fly. 

The finiteness of FDCA allows the definition of its complexity measure in terms of the complexity of realization of each cell. It is shown that a more detailed analysis of the complexity measure, than the one that involves input size only, is necessary to account for potential reduction of complexity via multilayered processing. Such analysis is done with the use of the cardinality of the class of local configurations recognized which is treated as a characteristic of the partitioning of all possible local configurations that is induced by the local map. The conditions for the two-layered FDCA to be preferred over one-layered FDCA are derived. These conditions relate sizes of the basic neighborhoods, the size of the input configurations, the cardinality of the class of local configurations recognized (section~\ref{sec:conditions}). Finally, a model of a common image filtering task satisfying these conditions is presented in section~\ref{sec:example}.

\section{Definitions}
\subsection{Finite dynamic cellular automata}
\label{sec:FDCA}

The following definition of FDCA is based on the definition in~\cite{Makatchev1997,Makatchev1998}.

\begin{defn} 
We define a \textit{dynamic cellular automaton (DCA)} as an infinite set $\mathcal{A}$ of layers $\sigma_i$,
 $\mathcal{A}=\{\sigma_{0}, \sigma_{1}, \dots\}$,  where $\sigma_{i}=(d, k_{i}, N_{i}, f_{i})$. Here $d$
denotes the dimension of the infinite grid of cells $C=\mathbb{Z}^d$. Every cell of the layer 
$i$ can assume states from a finite set 
$\Sigma_{i}=\{0,1,\ldots, k_{i}-1\}$, $k_{i} \ge 2$ of possible states 
that is called the alphabet of the layer $i$ automaton. The function $f_{i}: \Sigma_{i-1}^{N_{i}}\rightarrow \Sigma_{i}$  is the  local map of the layer $i$ automaton. For the sake of simplicity 
we assume that the {\it basic neighborhood} $N_{i}$ of the local map of layer $i$, 
is of form $N_{i} = [-r_{i, 1}, r_{i, 1}]\times[-r_{i,2},r_{i,2}]
\times \cdots \times [-r_{i, d}, r_{i,d}] \subseteq \mathbb{Z}^{d}$, 
where $r_{i,j}\ge 1, j=1,\ldots, d$.
Local map and basic neighborhood of layer $0$ are not defined.

We call a map $X_{i}: C \rightarrow \Sigma_{i}$ from the set $C$ of all cells to 
the alphabet of layer $i$ a {\it configuration} of the respective layer
the dynamic cellular automaton and denote as $\boldsymbol{X_{i}}$ the space of 
all possible configurations of this layer. 
Local map $f_{i}$ induces the {\it global map} 
$F_{i}:\boldsymbol{X_{i-1}}\rightarrow\boldsymbol{X_{i}}$. For a particular 
configuration $X_{i}$ and a cell $c\in C$ we define the 
{\it local configuration} $x_{i, c}$ at $c$ as the map 
$N_{i+1}\rightarrow\Sigma_{i}$, such that $x_{i, c}(z)=X_{i}(c+z)$. 
Now we can define the global map $F_{i}$ as $F_{i}(X_{i-1})(c) = f_{i}(x_{i-1,c})$.
\end{defn}

For the purpose of recognition of classes of bounded configurations it is useful to consider a finite subset of the grid of cells $C$ of DCA.

\begin{defn}
A \textit{finite dynamic cellular automaton (FDCA)} is a DCA with a distinguished finite subset of cells $W \subset C$ that is called the working zone of the FDCA. Without loss of generality we shall assume that $W=[w_{1}]\times[w_{2}]\times\cdots\times[w_{d}]$, where $[w]$ is the set $\{1, 2, \ldots, w\}$.
\end{defn}

Using both finite cellular automata and finite dynamic cellular 
automata one has to face the following technical difficulty: 
the local map cannot be 
applied to cells close to boundary of the working zone. There are known a few 
possible solutions to this problem for finite cellular automata. One 
of the solutions is to use {\it cyclic boundary conditions}, i. e.
glue together opposite ends of the working zone. The approach  used in this paper is to adopt {\it fixed 
boundary conditions} and assume that the missing cells are in some 
special state. 

\subsection{Recognition}
\label{sec:recognition}

\begin{defn}
We shall say that a FDCA $\mathcal{A}$ \textit{recognizes  a class of input configurations} $D \subseteq \boldsymbol{X_{0}}$ by the cell $c \in W$ of layer $i$, when there is a state $k \in \Sigma_{i}$ of the cell $c$ such that
\begin{displaymath}
X_{i}(c) = k \ \textrm{iff} \  X_{0} \in D.
\end{displaymath}

Similarly when referring to a cell $c$ of layer $i \ge 1$ of a FDCA we can talk about \textit{recognition  of a class of local configurations} $E \subseteq N_{i}^{\Sigma_{i-1}}$ of the preceding layer $i-1$, when there is a state $k \in \Sigma_{i}$ such that
\begin{displaymath}
X_{i}(c) \equiv f_{i}(x_{i-1,c}) = k \ \textrm{iff} \  x_{i-1,c} \in E.
\end{displaymath}
\end{defn}

%%%%24/4/02 Add def of recognition by a cell.

\subsection{Special cases}

Further we restrict our attention to two special cases of FDCA: 1-layered and 2-layered FDCA.

\begin{defn}
We shall call a FDCA $\mathcal{A}$ a \textit{1-layered FDCA} if there is a cell $c \in W$ such that 
$c + N_{1} = W$, where $N_{1}$ is the basic neighborhood of layer $1$. 
Similarly, we call a FDCA $\mathcal{A}$ a \textit{2-layered FDCA} if there is a cell $c \in W$ such that 
$c + N_{2} = W$, where $N_{2}$ is the basic neighborhood of layer $2$.
\end{defn}

The conditions below further restrict the types of FDCA investigated in this paper:
\begin{itemize}
\item dimension $d$ of the FDCA is 2,
\item basic neighborhoods are squares: $N_{i} = {[-r_{i}, r_{i}]}^2$,
\item working zone is a square:  $W = [w]^2$,
\item all alphabets are binary: $\Sigma_{i}=\{0,1\}$.
\end{itemize}

\begin{rem}
Since at the top layer of a FDCA there is at most one cell participating in recognition of a given class of configurations, we set its local neighborhood equal to the working zone $W$ and ignore the rest of the cells of this layer. 
\label{rem:topneighb}
\end{rem}

%Two typical 1-layered and 2-layered FDCA are shown in Fig.~\ref{fig:2FDCA}.
%\begin{figure}[htbp]
%\begin{center}
%\mbox{
%	\subfigure[1-layered FDCA]
%	{\includegraphics{1layered.eps}}\qquad
%	\subfigure[2-layered FDCA]
%	{\includegraphics{2layered.eps}}
%}%end mbox
%\caption{Special FDCA}
%\label{fig:2FDCA}
%\end{center}
%\end{figure}

\subsection{Complexity}
\label{sec:complexity}

\subsubsection{Cell complexity}

We define a complexity of  recognition of a class of input configurations $D$ by a FDCA $\mathcal{A}$  in terms of the complexity of its individual cells.

\begin{defn}
\label{def:cellcomplexity}
We denote the \textit{complexity of a cell} $c \in W$ of layer $i$ as a function $L(n, p)$, where $n$ is the size of the basic neighborhood, $n = |N_{i}|$, and $p$ is the cardinality of the class $x'$ of local configurations recognized by the cell $c$,  $p=|x'|$, $x' = \{ x_{i-1, c}: f_{i}(x_{i-1,c}) = 1 \}$. 
\end{defn}

\paragraph*{Assumption} The strong assumption we make about the complexity function $L(n, p)$ is that it is monotonically increasing with respect to $n$ and $p$. While it is natural to assume monotonicity of a complexity measure on the size of the input $n$, many real-life complexity measures are not monotone with respect to the cardinality of a partition class $p$ induced by a map on its domain. For example the length of the canonical DNF (disjunctive normal form) is monotone w.r.t. $p$, while the minimum of the lengths of canonical DNF and canonical CNF (conjunctive normal form) is not monotone w.r.t. $p$ \cite{wegener,yablonsky}. However, for many complexity measures (including the minimum of the lengths of canonical DNF and canonical CNF) it is possible to specify the reasonable range of $p$ for which this measure is monotone. Further we will assume that $p$ is always within the range of monotonicity of $L(n, p)$.

\subsubsection{Complexity of FDCA}

Consider a 1-layered FDCA $\mathcal{A}_1$  with the working area $W$ that recognizes a class of input configurations $D$ by a layer $1$ cell $c$. Let $N$ be the size of the input for a cell of layer $1$, then $N \equiv |N_1| = |W|$ due to remark \ref{rem:topneighb}. Let $P_1$ be the cardinality of the class of configurations recognized by a cell of layer $1$,  $P_1=|D|$.

From the definition of cell complexity (definition~\ref{def:cellcomplexity}), assuming that the cells within the same layer of FDCA run sequentially and ignoring the cells that do not participate in recognition (all but one cell of layer $1$), we can derive the expression for the \textit{sequential time complexity} of $\mathcal{A}_1$:
\begin{equation}
\lambda(\mathcal{A}_1) = L(N, P_1).
\label{eq:def1seqcompl}
\end{equation}

Since for a 1-layered FDCA there is only one cell that participates in the recognition, the \textit{parallel time complexity} of $\mathcal{A}_1$ is equal to its sequential time complexity:
\begin{equation}
\mu(\mathcal{A}_1) = L(N, P_1).
\label{eq:def1paralcompl}
\end{equation}

Similarly, we can derive the expressions for the sequential and parallel time complexities of a 2-layered FDCA.
Consider a 2-layered FDCA $\mathcal{A}_2$  with the working area $W$ that recognizes a class of input configurations $D$ by the layer $2$ cell $c$. Let $N$ be the size of the input for a cell of the layer $2$, then $N \equiv |N_2| = |W|$ due to remark \ref{rem:topneighb}. Let $P_2$ be the cardinality of the class of local configurations of layer $1$ recognized by the cell $c$ of layer $2$, $P_2=|\{x_{1,c}: f_{2}(x_{1,c}) = 1\}|$, let $p$ be the cardinality of the class of local configurations of layer $0$ recognized by a cell of layer $1$, $p=|\{x_{0,c'}: f_{1}(x_{0,c'}) = 1,\ c'\in W\}|$, and let $n$ be the size of the input for a cell of layer $1$, $n=|N_1|$. 

We can express the \textit{sequential time complexity} of $\mathcal{A}_2$ as
\begin{equation}
\lambda(\mathcal{A}_2) = N\cdot L(n,p)+L(N, P_2).
\label{eq:def2seqcompl}
\end{equation}

Assuming that the cells of the same layer run in parallel, the \textit{parallel time complexity} of $\mathcal{A}_2$ is 
\begin{equation}
\mu(\mathcal{A}_2) =  L(n,p)+L(N, P_2).
\label{eq:def2paralcompl}
\end{equation}

\subsection{Speedup}
\label{sec:speedup}

\begin{defn}
We shall say that the recognition of a class of configurations $D$ allows speedup via decomposition with respect to the cell complexity measure $L$ and sequential time complexity $\lambda$, if there exist a 1-layered FDCA $\mathcal{A}_1$ and a 2-layered FDCA $\mathcal{A}_2$ that both recognize $D$ and 
\begin{equation}
\frac{\lambda(\mathcal{A}_2)}{\lambda(\mathcal{A}_1)}\longrightarrow 0, \quad N \longrightarrow \infty.
\label{eq:defspeedup}
\end{equation}

Automaton $\mathcal{A}_2$ is then said to achieve the speedup over $\mathcal{A}_1$.
\end{defn}

Similarly, we can define the speedup with respect to cell complexity measure $L$ and the parallel time complexity $\mu$.

\section{Conditions for speedup}
\label{sec:conditions}

In this paper we shall concern ourselves with the speedup w.r.t. sequential time complexity $\lambda$.

\begin{thm}
\label{prop:criteria}
In the notation of sections \ref{sec:complexity} and \ref{sec:speedup}, FDCA $\mathcal{A}_2$ achieves the speedup over FDCA $\mathcal{A}_1$ (w.r.t. $\lambda$-complexity) in recognition of class of configurations $D$  iff the following two conditions are satisfied ($N \longrightarrow \infty$):
\begin{eqnarray}
\textrm{C1}:\qquad& &\frac{L(N, P_2)}{L(N,P_1)} \longrightarrow 0,\\
\textrm{C2}:\qquad& &\frac{N \cdot L(n, p)}{L(N,P_1)} \longrightarrow 0.
\end{eqnarray}
\end{thm}
\begin{pf}
Directly follows from (\ref{eq:def1seqcompl}), (\ref{eq:def2seqcompl}), (\ref{eq:defspeedup}) and the fact that $L(y, z) > 0$ provided $z>0$, and $P_1$, $P_2$, $p$ are all positive (this holds for nontrivial classes of configurations). \qed
\end{pf}

\begin{rem}
Condition~C1 immediately implies that the cell complexity measure $L(y, z)$ should essentially depend on its second argument $z$, the cardinality of the class of configurations recognized by the cell at a particular layer. 
\end{rem}

\begin{cor} 
Let $L(y, z) = \Theta(z^a)$ for some $a>0$, $L(y, z)$ is monotonically increasing w.r.t. $y$, $z$, then\\ (a) C1 is satisfied iff
\begin{equation}
\frac{P_2}{P_1} \longrightarrow 0,
\end{equation}
(b) for C2 to be satisfied, it is sufficient that
\begin{equation}
p = o\left(\frac{P_1}{N^{1/a}}\right).
\end{equation}
\end{cor}
\begin{pf}
(a) Immediately follows from the asymptotics $L(y, z) = \Theta(z^a)$ and the definition of C1.

(b) Since $L(y, z)$ is monotone on $y$, 
\begin{displaymath}
\frac{L(N, P_1)}{N\cdot{}L(N,p)}\longrightarrow\infty \Rightarrow \frac{L(N, P_1)}{N\cdot{}L(n,p)}\longrightarrow\infty.
\end{displaymath}
${L(N, P_1)}/\left({N\cdot{}L(N,p)}\right) \rightarrow \infty$ $\Leftrightarrow$ ${{P_1}^a}/(N p^a) \rightarrow \infty$ $\Leftrightarrow$ $p = o\left({P_1}/{N^{1/a}}\right)$. \qed
\end{pf}

\section{Example of speedup}
\label{sec:example}

In this section we design a class of configurations and respective 1- and 2-layered FDCA recognizing the class that satisfy conditions C1 and C2 for the speedup. 

\begin{figure}[tbp]
\begin{center}
\mbox{
	{\includegraphics{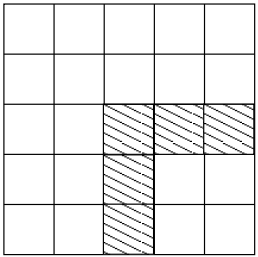}}
}%end mbox
\caption{A local configuration corresponding to the angular pattern in Definition~\ref{def:angle} (Shaded cells are in state $1$, the others are in state $0$).}
\label{fig:examplepattern}
\end{center}
\end{figure}

\begin{defn}
\label{def:angle}
Consider a particular type of angular pattern on the square grid of area $n$ such that $\sqrt{n}$ is odd. An example of the corresponding local configuration $\bar{x}$ of area $n = 25$ is shown in figure~\ref{fig:examplepattern}. Denote as $x'$
the class of all local configurations within the Hamming distance of $1$ from $\bar{x}$, $x' = \{x: d(x, \bar{x}) \le 1)\}$. For a working zone $W$ of a size $N > n$ we shall define a class of input configurations $X'$, $X' = \{X: X\ \textrm{includes exactly one instance of a pattern from}\ x'\}$. Let us denote quantities $|x'| = n+1$ as $p$,  and $|X'|$ as $P_1$.
\end{defn}

\begin{rem} Informally, class $X'$ consists of all the configurations that have a single occurrence of the angular pattern that is specified with the accuracy of up to one possible error (thus the Hamming distance of $1$).
\end{rem}

Further we shall be concerned with the problem of recognition of the class of input configurations $X'$. Let us specify two FDCA recognizing $X'$.

We shall denote the one-layered FDCA that recognizes $X'$ as $\mathcal{A}_1$. As is easy to see 
\begin{equation}
	\lambda(\mathcal{A}_1) = L(N, P_1).
\end{equation}

The two-layered FDCA  $\mathcal{A}_2$ that recognizes $X'$ is defined so that cells of its first layer recognize $x'$ via the local map from the basic neighborhood of size $n$ and the cell of its second layer recognizes the class of configurations $X''= \{X : X\ \textrm{has exactly one cell in state $1$}\}$. If $B$ is the number of cells of layer $2$ that fall under the fixed boundary condition mentioned in ~\ref{sec:FDCA}, then $|X''|= N - B$ --- the quantity that we will denote as $P_2$.  
Note that
\begin{equation}
\lambda(\mathcal{A}_2) = N\cdot L(n,p)+L(N, P_2).
\label{eq:A2comp}
\end{equation}

\begin{prop}
Let $L(y, z) = \Theta(z^a)$ for some $a>0$, $L(y, z)$ is monotonically increasing w.r.t. $y$, $z$, and $n = O(N^{1/3a})$. The recognition of the class of configurations $X'$ allows speedup via decomposition.  
\end{prop}
\begin{pf}
To prove the statement of the theorem it is sufficient to show that $\mathcal{A}_1$ and  $\mathcal{A}_2$ realize the speedup. This will be done via Theorem~\ref{prop:criteria}, namely we shall show that $\mathcal{A}_1$ and  $\mathcal{A}_2$ satisfy conditions C1 and C2. 

Condition C1 is satisfied if ${P_2}/{P_1} \longrightarrow 0$. To prove the latter limit, let us construct a superlinear lower bound on size of ${P_1}$, bearing in mind that $P_2 = N-B$. A simple estimate of number of possible configurations in $X'$ yields
\begin{equation}
\label{eq:P1estim2}
P_1 > (N-B) (n+1) (N-n-2)^3,
\end{equation}
which is sufficient for our purpose.

To prove C2  using the property of monotonicity of function $L(y, z)$ it is sufficient to show that  
\begin{equation}
\label{eq:proofC2-2}
\frac{Np^a}{{P_1}^a} \longrightarrow 0,
\end{equation}
which follows from (\ref{eq:P1estim2}), $p = n+1$ and the condition of the theorem.
\qed
\end{pf}

\section{Conclusion}
We derived the conditions for a 2-layered FDCA to be more efficient (w.r.t the sequential time complexity measure) than a 1-layered FDCA, under the assumption that the cell complexity is polynomial on the cardinality of the class of configurations being recognized. This assumption holds for some complexity measures, such as the length of the canonical DNF, and it has plausibility regions for others, such as minimum of the lengths of the canonical DNF and CNF.  A real-life image pattern recognition problem is formalized and is shown to allow the speedup via utilization of a 2-layered FDCA.

Among the interesting problems that remain to be solved we mention improvement of the sufficient conditions of the speedup, and investigating speedup with respect to other plausible cell complexity measures.

\section{Acknowledgments}
The approach to studying FDCA and their complexity presented in this paper were introduced in the thesis \cite{Makatchev1997}, written in the inspirational environment of the Chair of Mathematical Cybernetics, Moscow State University, under supervision of V.~N.~Kozlov to whom the author is greatly indebted. The author would like to thank A.~P.~Ryjov for his constructive critique of the thesis.  O.~P.~Skobtsov supported the work in numerous ways. The author is currently supported by Why2-Atlas project at CIRCLE/LRDC, MURI grant N00014-00-1-0600 from ONR Cognitive Science and by NSF grant 9720359. 

\

\end{document}